\documentclass[twocolumn]{IEEEtran}

\usepackage{amsfonts}
\usepackage{amssymb}
\usepackage{graphicx}
\begin{document}
			   
\title{Shot noise suppression in quasi-one dimensional Field Effect Transistors}
\author{Alessandro Betti, Gianluca Fiori and Giuseppe Iannaccone\\
Dipartimento di Ingegneria dell'Informazione: Elettronica, Informatica, Telecomunicazioni, \\
via Caruso 16, 56100 Pisa, Italy\\
email: {\{alessandro.betti, g.fiori, g.iannaccone\}@iet.unipi.it}, Tel. +39 050 2217639}

\maketitle

\bibliographystyle{unsrt}



\begin{abstract}

We present a novel method for the evaluation of shot noise in 
quasi one-dimensional field-effect transistors, such as those based
on carbon nanotubes and silicon nanowires. The method is derived by using
a statistical approach within the second quantization formalism and allows to 
include both the effects of Pauli exclusion and Coulomb repulsion among charge 
carriers. In this way it extends Landauer-B\"uttiker approach by explicitly 
including the effect of Coulomb repulsion on noise. We implement the method
through the self-consistent solution of the 3D Poisson and transport equations 
within the non-equilibrium Green's function framework and a Monte 
Carlo procedure for populating injected electron states. 
We show that the combined effect of Pauli and Coulomb interactions reduces 
shot noise in strong inversion down to 23\% of the full shot 
noise for a gate overdrive of 
0.4~V, and that neglecting the effect of Coulomb repulsion would lead to an 
overestimation of noise up to 180\%. 
\end{abstract}


\IEEEpeerreviewmaketitle
\IEEEoverridecommandlockouts

\section{Introduction}

In the last few years, a huge collective effort has been directed to 
assess potential performance of quasi-1D Field Effect Transistors (FETs) 
based on 
Carbon Nanotubes~\cite{Martel,JGuo,GKlimeck} (CNTs), 
Silicon NanoWires~\cite{YCui} (SNWs), Graphene nanoribbons (GNRs) versus the 
International Technology Roadmap for Semiconductors~\cite{ITRS} (ITRS) 
requirements, both from an experimental and a theoretical point of view. 
However, attention has been focused on electrical quantities like 
$I_{\rm on}/I_{\rm off}$, subthreshold slope, mobility, transconductance~\cite{Yuzvinsky,VPerebeinos,JKnoch}, while an 
accurate investigation of electrical noise has been often neglected. 
Although the 1/f noise represents the major noise source 
affecting CNT-FETs performance~\cite{YMLin,JAppenzeller}, 
the intrinsic shot noise is not 
only critical from an analog and digital design point of view, 
but can also provide relevant information regarding interactions 
among carriers~\cite{ik1prb}~\cite{ik1}~\cite{BlanterBut}, 
electron energy distribution~\cite{Gramespacher}~\cite{Bul2} and 
electron kinetics~\cite{Land}. 

Due to the limited device dimensions, even in strong inversion only 
few electrons take part to transport, so that drain current fluctuations 
can heavily affect device electrical behavior. 
Pauli and Coulomb interactions play an important role in 
noise analysis, through fluctuations of the occupation number of 
injected states and fluctuations of the 
potential barrier along the channel. 

From a numerical point of view, 
a self-consistent solution of the electrostatics and transport 
equations is mandatory in order to properly consider such effects.  
An analysis of this kind has been performed for example in double gate 
MOSFETs~\cite{YNaveh} and in nanoscale ballistic MOSFETs~\cite{ik1_2}, 
where a strong shot noise suppression, mostly due to Pauli exclusion 
principle, has been observed.

A different approach, based on quantum trajectories within the De 
Broglie-Bohm framework has been instead presented in~\cite{XOriols1},
where resonant tunneling diodes have been studied and heavy 
approximations have been adopted in order to easily consider electron-electron 
correlation in the many-body problem. 

Actually, a complete understanding of the mechanism of suppression of shot 
noise in CNT and SNW-FETs is still a debated issue. 
Indeed, the significant suppression of current fluctuations by more 
than a factor 100 
obtained at low temperature for suspended ropes 0.4 $\mu$m long of 
single wall carbon nanotubes~\cite{PRoche} has not been supported by a 
comprehensive theoretical analysis. 
Recent experiments of shot noise in CNT-based Fabry Perot 
interferometers~\cite{Hermann} show that by only including Pauli exclusion one 
is able to explain most of the dependence of shot noise on the backgate bias, 
but in some bias conditions additional mechanisms of electron-electron 
interaction might be needed to explain the observed noise suppression. 
Theoretical  efforts have been mainly addressed to 
model the electrical noise in SNW-FETs, where, within a scattering approach 
with the limitation of excluding space-charge effects 
on electron transmission, Pauli exclusion reduces electrical noise 
in strong inversion down to 0.6\% of the full value for a gate overdrive of 
0.3~V~\cite{Park1}, whereas an interesting 
increase of noise is observed 
by including electron-phonon scattering processes~\cite{Park2}. 

Here, we present a new method to compute the shot noise power 
spectral density in ballistic CNT and SNW-FETs based on Monte Carlo (MC) 
simulations of randomly injected electrons from the reservoirs. 
In order to consider correlations between fermions, an analytical formula 
for the noise power spectral density has been 
computed by means of a statistical approach within the second quantization 
formalism. The derived formula has then been 
implemented in the self-consistent solution of the 3D Poisson and Schr\"odinger equations, within the NEGF formalism. 

\section{Theory}  

The average current in a mesoscopic conductor can 
be expressed by means of Landauer's formula:
\begin{equation} \label{eqn:Landauer}
\langle I \rangle= \frac{e}{\pi\hbar} \int d\!E \left\{ \mathrm{Tr}\left[\mathbf{t^\dagger t}(E)\right]\left[f_S(E)-f_D(E)\right] \right\}
\end{equation}
where $\mathbf{t}$ is the transmission amplitude matrix for states emitted 
from the source (S) and collected at the drain (D) and $f_S$ 
and $f_D$ are the Fermi-Dirac statistics of the S and D, respectively.

The zero-frequency noise power spectral density for a 
two-terminal conductor - the so-called Landauer-B\"uttiker noise formula - reads~\cite{MBut2}~\cite{TMartin}:
\begin{eqnarray} \label{eqn:noise}
S(0) \!\! \! \! \! \! &=& \! \!\!\! \!  \!\frac{2\,e^2}{\pi\hbar}\!\! \int\! \!d\!E \left\{ \left[f_S (1\!-\!f_S)+f_D (1\!-\!f_D)\right]\mathrm{Tr}\!\left[\mathbf{t^\dagger t t^\dagger t}\right] \right. \nonumber \\
&+& \!\!\!\!\!\!\left.  \left[f_S (1\!-\!f_D)+f_D (1\!-\!f_S )\right] \!\left( \mathrm{Tr}\!\left[\mathbf{t^\dagger t} \right]\! -\!\mathrm{Tr}\!\left[\mathbf{t^\dagger t t^\dagger t} \right]\right)\! \right\} \, , \nonumber \\
\end{eqnarray}
where $\mathbf{t^\dagger}$ is the conjugate transpose of the matrix 
$\mathbf{t}$. 
However, eq.~(\ref{eqn:noise}) holds only if one assumes that fluctuations of 
the potential profile do not occur, i.e. that Coulomb interaction between 
carriers is completely neglected.
Actually, the potential barrier along the channel fluctuates in time, since 
randomly injected electrons modify the height of the barrier through 
long-range Coulomb interaction, which in turn affects carriers transmission 
and eventually leads to the suppression of the drain current fluctuations.  

In order to compute the expression of the power spectral density in the 
general case, we take 
advantage of the second quantization formalism.
In particular, at zero magnetic field, the time-dependent current operator at 
the source can be expressed as the difference between 
the occupation numbers of 
carriers moving inward and outward the source contact in each quantum channel~\cite{MBut2} ($n_{Sm}^+$ and $n_{Sm}^-$, respectively):
\begin{eqnarray} \label{eqn:current}
I(t) \!\! \! \! &=& \! \!\! \!\frac{e}{2 \pi\hbar} \sum_{m \in S} \int d\!E \left[n_{Sm}^+(E,t) - n_{Sm}^-(E,t)\right] \, ,
\end{eqnarray} 
where 
\begin{eqnarray} \label{eqn:n+}
n_{Sm}^+(E,t) \!\! \! \! &=& \! \!\! \!\int d(\hbar \omega) a_{Sm}^+(E)\,a_{Sm}(E+\hbar \omega)\, e^{-i\omega t} \, , \nonumber \\
n_{Sm}^-(E,t) \!\! \! \! &=& \! \!\! \!\int d(\hbar \omega) b_{Sm}^+(E)\,b_{Sm}(E+\hbar \omega)\, e^{-i\omega t} \, .
\end{eqnarray}
The operators $a_{Sm}^{\dagger}(E)$ and $a_{Sm}(E)$ create and 
annihilate, respectively, incident electrons in the source lead with total 
energy $E$ in the transverse channel $m$. 
In the same way, the creation $b_{Sm}^{\dagger}(E)$ and 
annihilation $b_{Sm}(E)$ operators refer to electrons in the source lead 
for outgoing states. 
For the CNT case, the channel index $m$ runs over all the transverse modes and 
different spin, whereas for SNW, it also runs along the 
six minima of the conduction band in the $\mathbf{k}$ space.
In addition, the operators $a_S$ and $b_S$ are related through the 
unitary transformation
\begin{eqnarray} \label{eqn:relationab}
b_{Sm}(E)=\sum_{\alpha = S,D}
\sum_{n \in \alpha }^{N_{\alpha}}
\mathbf{s}_{S\alpha;mn}(E)a_{\alpha n}(E) \, ,
\end{eqnarray}
where the scattering matrix $\mathbf{s}$ has dimensions $(N_S + N_D)\times (N_S + N_D)$ 
and $N_S$ and $N_D$ are the number of quantum channels in the source and drain 
contacts, respectively. In the following, time dependence will be neglected, 
since we are interested to the zero frequency case.

If $\mid \! \sigma \rangle$ is a many-particle (antisymmetrical) 
state, 
the occupation number $\sigma_{\alpha m}(E) $ in the reservoir 
$\alpha$ ($\alpha= S,D $) in the channel $m$ can be either 0 or 1, 
and can be 
expressed as $\sigma_{\alpha m}(E)= \langle a_{\alpha m}^{\dagger}(E)a_{\alpha m}(E)\rangle_{\sigma}$.
Since we are interested to current fluctuations, we need to consider 
an ensemble of many electrons states $\{ \mid \sigma_1 \rangle,\mid \sigma_2 \rangle,\mid \sigma_3 \rangle,\cdots,\mid \sigma_N \rangle \}$ and to compute 
statistical averages $\langle \, \rangle_s$. 
By assuming no correlations between states at different energy or injected 
from different reservoirs, the statistical average of $\sigma_{\alpha m}(E)$ 
reads 
\begin{eqnarray} \label{eqn:sigma}
\langle \sigma_{\alpha m}(E) \rangle_s= \,\langle \langle a_{\alpha m}^{\dagger}(E) a_{\alpha m}(E)\rangle_{\sigma}\rangle_s = \, f_{\alpha}(E) 
\end{eqnarray}
In the following, we identify $\langle \langle \,\,\, \rangle_{\sigma}\rangle_{s}$ with $\langle \,\,\, \rangle$.
By means of (\ref{eqn:relationab}), we obtain the mean current:
\begin{eqnarray} \label{eqn:meancurrent}
\langle I \rangle \!\!\!\!&=&\!\!\!\! \frac{e}{2 \pi \hbar} \int d\!E \,\left\{\sum_{n \in S} \langle \left[\mathbf{t^{\dagger} t} (E) \right]_{nn} \sigma_{S n}(E) \rangle_s \right. \nonumber \\
&-&\!\!\! \left. \sum_{k \in D} \langle \left[\mathbf{t'^{\dagger}t'}(E) \right]_{kk} \sigma_{D k} (E) \rangle_s \right\}
\end{eqnarray}
where $ \mathbf{t'}$ is the drain-to-source transmission amplitude matrix~\cite{Datta} .
Since $ \sigma_{\alpha m}^2= \sigma_{\alpha m} $, $\forall m \in \alpha$ and exploiting the unitarity of the scattering matrix, the 
mean squared current fluctuation for unit of energy can be expressed as:
\begin{eqnarray} \label{eqn:variance}
\frac{var\left(I \right)}{\Delta E}\!\!\!\!\!\! &=&\!\!\!\!\!\! \left(\frac{e}{h}\right)^2 \!\!\!\int \!\! d\!E \!\!\! \sum_{\alpha = S,D} \sum_{l \in \alpha} \langle  \left[\mathbf{\tilde{t}}\right]_{\alpha;ll}\left(1-\left[\mathbf{\tilde{t}}\right]_{\alpha;ll}\right) \! \sigma_{\alpha l}\rangle_s \nonumber \\
&-&\!\!\!\! \left(\frac{e}{h}\right)^2 \! \!\!\int \! d\!E \!\!\! \sum_{\alpha = S,D} \!\!\!\!
\sum_{
\begin{array}{c}
\scriptstyle l,p \in \alpha \\
\scriptstyle l \neq p \\
\end{array}} 
\!\!\!\!\langle \left[\mathbf{\tilde{t}}\right]_{\alpha;l p} \left[\mathbf{\tilde{t}}\right]_{\alpha;p l} \sigma_{\alpha l} \sigma_{\alpha p}\rangle_s \nonumber \\
&-& \!\!\!\!2  \left(\frac{e}{h}\right)^2 \!\!\!\!\int \! \!d\!E  \! \sum_{k \in D} \sum_{p \in S}\! \langle \left[\mathbf{t'^{\dagger}r}\right]_{kp} \left[\mathbf{r^{\dagger}t'}\right]_{pk}  \sigma_{Dk} \sigma_{Sp} \rangle_s  \nonumber \\
&+& \!\!\!\!\!\!\frac{1}{\Delta E} var \! \left\{\! \!\frac{e}{h}\! \int \!\!\!d\!E \!\left(\sum_{n \in S} \!\!\left[\mathbf{\tilde{t}} \right]_{S;nn}\! \sigma_{S n}  \! 
- \!\!\! \sum_{k \in D}  \!\!\left[\mathbf{\tilde{t}} \right]_{D;kk}\! \sigma_{D k} \!\! \right)\!\!\right\} \nonumber \\ 
\end{eqnarray} 
where $\left[\mathbf{\tilde{t}}\right]$ is defined as
$$
\left[\mathbf{\tilde{t}}\right]_{\alpha;lp}=
\left\{
\begin{array}{cc}
 \left[\mathbf{t^{\dagger} t}\right]_{lp}  \mbox{ \quad  if  } \alpha=S\\
  \left[\mathbf{t'^{\dagger} t'}\right]_{lp}                        \mbox{ \quad  if  } \alpha=D  \, ,
\end{array}
\right.
$$
and $\mathbf{r}$ is the reflection amplitude matrix~\cite{Datta}. 
$\Delta E$ is our energy step of choice, i.e. the minimum energy separation 
between injected states. 

Eq.~(\ref{eqn:variance}) is expressed as the sum of four terms: 
the first, the second and the third terms correspond to the partition 
noise contribution. In particular, the first term is strictly related on 
the quantum uncertainty of the transmission process and disappears in the 
classical limit; the second term is associated to the correlation between 
transmitted states coming from the same reservoir; the third term 
to the correlation between transmitted and reflected states in the source 
lead; the minus sign in the second and third terms is due to exchange 
pairings, because of the fermionic nature of the electrons. 
In particular, the second and the third terms provide physical insights 
on exchange interference effects~\cite{ABetti2}.
Finally, the last term represents the injection noise obtained as the 
variance computed on the ensemble of current samples.

According to the \begin{em}Milatz Theorem\end{em}~\cite{Ziel}, the noise 
power spectral density in the zero frequency limit can be computed as 
$ S(0)= lim_{f \rightarrow  0}\, S(f) = lim_{\nu \rightarrow  0} \left[2/\nu  \cdot var(I)\right] $, 
where  $ \nu $ is the injection rate, which can be expressed as: 
\begin{eqnarray} \label{eqn:nu}
 \nu = \Delta E/(2 \pi \hbar) 
.
\end{eqnarray} 
Eventually, the power spectral density of shot noise at zero frequency can be 
expressed as:
\begin{equation} \label{eqn:noisepower}
S(0) = \lim_{\nu \rightarrow  0}\frac{2}{\nu} \, var(I)= \lim_{\Delta E \rightarrow  0} 4 \pi \hbar \frac{var(I)}{\Delta E} 
\end{equation}
It is worth noticing that eqs.~(\ref{eqn:noisepower}) 
and~(\ref{eqn:variance}) are not equivalent to the Landauer-B\"uttiker's 
formula~(\ref{eqn:noise}), since in eq.~(\ref{eqn:variance}) 
the transmission ($\mathbf{t}$, $\mathbf{t'}$) and 
reflection ($\mathbf{r}$) matrix are expressed as functionals of the 
statistics of the occupation of injected states 
from both contacts. In this way we are able to consider the fluctuation 
in time of the conduction and valence 
band edge profiles produced by the random injection through 
long-range Coulomb repulsion, providing a further source of noise suppression 
not included in eq.~(\ref{eqn:noise}). 

Indeed, from an analitical point of view, eqs.~(\ref{eqn:variance}) 
and~(\ref{eqn:noisepower}) reduce to eq.~(\ref{eqn:noise}) when 
transmission and reflection do not depend, through Coulomb interaction, 
on random occupation numbers of injected states: in that case we can take 
the terms related to transmission and reflection out of the statistical 
averages in~(\ref{eqn:variance}). 
By means of ~(\ref{eqn:sigma}) and exploiting $ \langle \sigma_{\alpha l}(E) \sigma_{\beta n}(E')\rangle_s = f_{\alpha}(E)f_{\beta}(E')+ \delta(E-E')\delta_{\alpha \beta} \delta_{l n} [f_{\alpha}(E)$ $-f_{\alpha}(E)f_{\beta}(E')]$, 
the fourth term in~(\ref{eqn:variance}) becomes:
\begin{eqnarray} \label{eqn:term4}
&&\left(\frac{e}{h}\right)^2\! \int \!d\!E \! \sum_{n \in S}  \left[\mathbf{t^{\dagger} t(E)} \right]_{nn}^2 f_{S}(E) \left[1-f_{S}(E)\right] \nonumber \\
&+&\!\!\!\left(\frac{e}{h}\right)^2\! \int \!d\!E \! \sum_{k \in D}  \left[\mathbf{t t^{\dagger}(E)} \right]_{kk}^2 f_{D}(E) \left[1-f_{D}(E)\right]
\end{eqnarray} 
since at zero magnetic field $\mathbf{t'^{\dagger}t'}=\mathbf{t t^{\dagger}}$. 
The terms $\delta(E-E')$, $\delta_{\alpha \beta}$ and $\delta_{l n}$ are 
the Kronecker delta. 
Taking advantage of $\sum_{k \in D} \sum_{p \in S} \left[\mathbf{t'^{\dagger} r}\right]_{kp} \left[\mathbf{r^{\dagger} t'} \right]_{pk}= \mathrm{Tr}\left[\mathbf{t^\dagger t} \right]-\mathrm{Tr}\left[\mathbf{t^\dagger t t^\dagger t} \right] $, $S(0)$ becomes:
\begin{eqnarray} \label{eqn:nonSClimit}
S(0) \!\! \!\!\!\! &=& \!\!\!\! \!\!  \lim_{\Delta E \rightarrow  0} 
4 \pi \hbar \, \frac{var(I)}{\Delta E} \nonumber \\
&=& \!\! \!\! \!\! \frac{2\,e^2}{\pi\hbar} \!\!\int \!\!d\!E \left\{ \left[f_S (1 \!- \!f_S)+f_D (1 \!- \!f_D)\right]\mathrm{Tr}\left[\mathbf{t^\dagger t t^\dagger t}\right] \right. \nonumber \\
&+& \!\!\!\!\!\!\left.  \left[f_S (1 \!\!- \!\!f_D)\!+\!f_D (1 \!\!- \!\!f_S )\right] \!\left( \mathrm{Tr}\left[\mathbf{t^\dagger t} \right]\!-\!\mathrm{Tr}\left[\mathbf{t^\dagger t t^\dagger t} \right]\right) \right\} \,  \nonumber \\
\end{eqnarray} 
which is Landauer-B\"uttiker's formula~(\ref{eqn:noise}).

Let us now point out that eq.~(\ref{eqn:noisepower}) would also 
simplify when identical and independent 
injected modes from the reservoirs are considered. In this case, 
$\mathbf{t}$, $\mathbf{t'}$ and $\mathbf{r}$ are all diagonal, so that the 
second term in~(\ref{eqn:variance}) becomes negligible. 
By exploiting the reversal time symmetry ($\mathbf{t'}= \, \mathbf{t^t}$, 
where $\mathbf{t^t}$ is the transpose of $\mathbf{t}$) 
and the unitarity of the scattering matrix, the power spectral 
density becomes: 
\begin{eqnarray} \label{eqn:noisesimulation}
S(0) \! \!\!\!\! \!  &=&  \!\!\!\!\!\!    \frac{e^2}{\pi\hbar} \! 
\left\{\! 
\int \!\!d\!E  \!\!\!\!\sum_{\alpha = S,D} \sum_{l \in \alpha} \langle \left[\mathbf{\tilde{t}}\right]_{\alpha;ll} \!\left(1\!-\!\left[\mathbf{\tilde{t}}\right]_{\alpha;ll}\right) \sigma_{\alpha l}\rangle_s \right. \nonumber \\
&-&\!\! 2 \int d\!E \sum_{l \in S}  \langle \left[\mathbf{\tilde{t}}\right]_{S;ll} 
\left(1- \left[\mathbf{\tilde{t}}\right]_{S;ll}\right) \sigma_{Dl} \sigma_{Sl} \rangle_s  
\nonumber \\
&+&\!\!\!\!\!\!  \frac{1}{\Delta E}\!\! \left. var \! \left[\!\int \!\!\!d\!E \! \left(\sum_{n \in S} \! \left[\mathbf{\tilde{t}}\right]_{S;nn} \sigma_{S n} \! - \!\!\! \sum_{k \in D} \! \left[\mathbf{\tilde{t}}\right]_{D;kk} \sigma_{D k}  \right)\!\right]\! \right\} \nonumber \\
\end{eqnarray}
  
\section{Simulation Methodology}
In order to properly include the effect of Coulomb interaction, we 
self-consistently solve the 3D Poisson equation imposing Dirichlet 
boundary conditions in correspondence of the metal gates, and 
null Neumann boundary conditions on the ungated surfaces which define the 3D 
domain. Within a self-consistent scheme, the 3D Poisson equation is coupled 
with the Schr\"odinger 
equation with open boundary conditions, within the Non-Equilibrium Green's 
function (NEGF) formalism which has been implemented in our in-house open 
source simulator {\sl NanoTCAD ViDES}~\cite{ViDES}. 
In particular the 3D Poisson equation reads 
\begin{eqnarray} \label{eqn:Poisson}
\nabla\left(\epsilon \nabla \phi\left(\vec{r}\right) \right)= -\left(\rho \left(\vec{r}\right)+\rho_{fix}\left(\vec{r}\right)\right) \, ,
\end{eqnarray} 
where $\phi$ is the electrostatic potential, $\rho_{fix}$ is 
the fixed charge which accounts ionized impurities in the doped regions, 
while $\rho$ is the charge density per unit volume
\begin{eqnarray} \label{eqn:density}
\rho \left(\vec{r}\right) \!\!\!&=&\!\!\! - e\!\int_{E_i}^{+\infty}\!\! \!d\!E \!\sum_{\alpha= S,D} \sum_{n \in \alpha} DOS_{\alpha n}\left(\vec{r},E\right) \sigma_{\alpha n}(E)  \nonumber \\ 
&+&\!\!\!e \!\!\int_{-\infty}^{E_i} \!\!\!\!\!d\!E \!\!\!\sum_{\alpha= S,D} \sum_{n \in \alpha}\! DOS_{\alpha n}\left(\vec{r},E\right)\! \left[1\!-\!\sigma_{\alpha n}(E)\right] ,
\end{eqnarray} 
where $E_i$ is the mid-gap potential, $DOS_{\alpha n}(\vec{r},E)$ is the 
local density of states associated to channel $n$ injected 
from contact $\alpha$ and $\vec{r}$ is the 3D spatial coordinate. 

From a numerical point of view, in order to 
model the stochastic injection of electrons from the contacts, 
a statistical simulation on an ensemble of random configurations 
of injected electron states from both contacts has been performed. 
In particular, we have uniformly discretized with step $\Delta$E 
the whole energy range of 
integration [equations~(\ref{eqn:variance}) and~(\ref{eqn:density})].
Each random injection configuration has been obtained by extracting a random 
number $r$ uniformly distributed between 0 and 1 for each state represented 
by energy $E$, reservoir $\alpha$, and quantum channel $n$. 
The state is occupied if $r$ is smaller than the Fermi Dirac factor, i.e. 
$\sigma_{Sn}(E)$ $[\sigma_{Dn}(E)]$ is 1 for 
$ r < f_S(E)\, [f_D(E)]$, and 0 otherwise. 

Self-consistent simulations for a given actual random statistics in the 
source and drain contacts have been then performed, 
and, at convergence, the transmission ($\mathbf{t}$, $\mathbf{t'}$) and 
reflection ($\mathbf{r}$) matrix have been computed, 
obtaining an element of the ensemble. 
In particular, for an actual electron distribution in the contacts, the 
Schr\"odinger equation is solved in order to obtain the spatial charge 
distribution~(\ref{eqn:density}) along the channel. 
Then, the latter is included in eq.~(\ref{eqn:Poisson}) and the electrostatic 
potential is then computed and, once convergence of the NEGF-Poisson 
iteration scheme is achieved, the scattering matrix is evaluated, and a 
new sample to be added to the noise ensemble is obtained. 
Finally, the power spectral density $S(0)$ can be extracted by means 
of eqs.~(\ref{eqn:variance}) and~(\ref{eqn:noisepower}).
From a computational point of view, we have verified that $S(0)$ 
computed on a record of 
500 samples, using the energy step $\Delta E= $~5~$\times$10$^{-4}$~eV, 
represents a good tradeoff between computational cost 
and accuracy of results~\cite{ABetti}.

Let us mention the fact that our approach is based on a mean field 
approximation of the Coulomb interaction, and that therefore the exchange term 
is not included. 
In the following, we will refer to self-consistent (SC) simulations 
when the Poisson-Schr\"odinger equations are solved considering 
$f_S$ and $f_D$ in 
eq.~(\ref{eqn:density}), while we refer to self-consistent Monte Carlo 
simulations (SC-MC), when statistical simulations with random occupations 
$\sigma_{Sn}(E)$ 
and $\sigma_{Dn}(E)$ inserted in eq.~(\ref{eqn:density}) are used. 
SC-MC simulations of randomly injected electrons allow to consider both the 
effect of Pauli and Coulomb interaction on noise. 
As a test, we have verified that if we perform MC simulations, keeping the 
potential profile along the 
channel fixed and exploiting the one obtained by means of SC simulation, the 
noise power spectrum computed in this way reduces to the 
Landauer-B\"uttiker's limit~(\ref{eqn:noise}), as already proved in an 
analitical way (eq.~(\ref{eqn:nonSClimit})): we refer to such simulations 
as non-self consistent Monte Carlo simulations (non SC-MC).
 
\section{Simulation Results} 
\subsection{Considered devices}
\begin{figure} [tbp]
\begin{center}
\includegraphics[width=8.6cm]{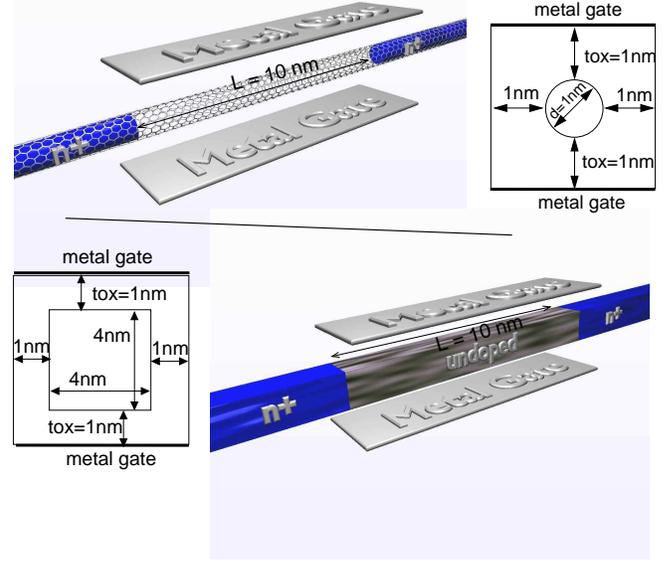}
\end{center}
\caption{3-D structures and transversal cross sections of the simulated CNT 
(top) and SNW-FETs (bottom).}
\label{fig:struttura}
\end{figure}

The simulated device structures are depicted in 
Fig.~\ref{fig:struttura}.
We consider a (13,0) CNT embedded in SiO$_{2}$ with 
oxide thickness equal to 1 nm, an 
undoped channel of 10 nm and n-doped CNT extensions 10 nm long, with a molar 
fraction $f=\,5 \times 10^{-3}$. 
The SNWT has an oxide thickness $t_{ox}$ equal to 1 nm and the 
channel length $L$ is 10 nm. The channel is undoped and the source and drain 
extensions (10 nm-long) are doped with $ N_D = \, 10^{20} $ cm$^{-3}$. The 
device cross section is 4$\times$4 nm$^2$.
A p$_z$-orbital tight-binding Hamiltonian has been assumed for 
CNTs~\cite{GFiori2,JGuo2}, whereas an effective mass 
approximation has been considered for SNWTs~\cite{GFiori1,JWang} by means of 
an adiabatic decoupling in a set of
two-dimensional equations in the transversal plane and in a set of 
one-dimensional equations in the longitudinal direction for each 1D subband. 

For both devices, we have 
developed a quantum fully ballistic transport model with semi-infinite 
extensions at their ends. 
A mode space approach has been adopted, since only 
the lowest subbands take part to transport: we have verified that 
four modes are enough to compute the mean current 
both in the ohmic and saturation region. 
All calculations have been performed at the temperature $T$= 300 K.

\subsection{DC Characteristics}
\begin{figure} [tbp]
\begin{center}
\includegraphics[width=8.6cm]{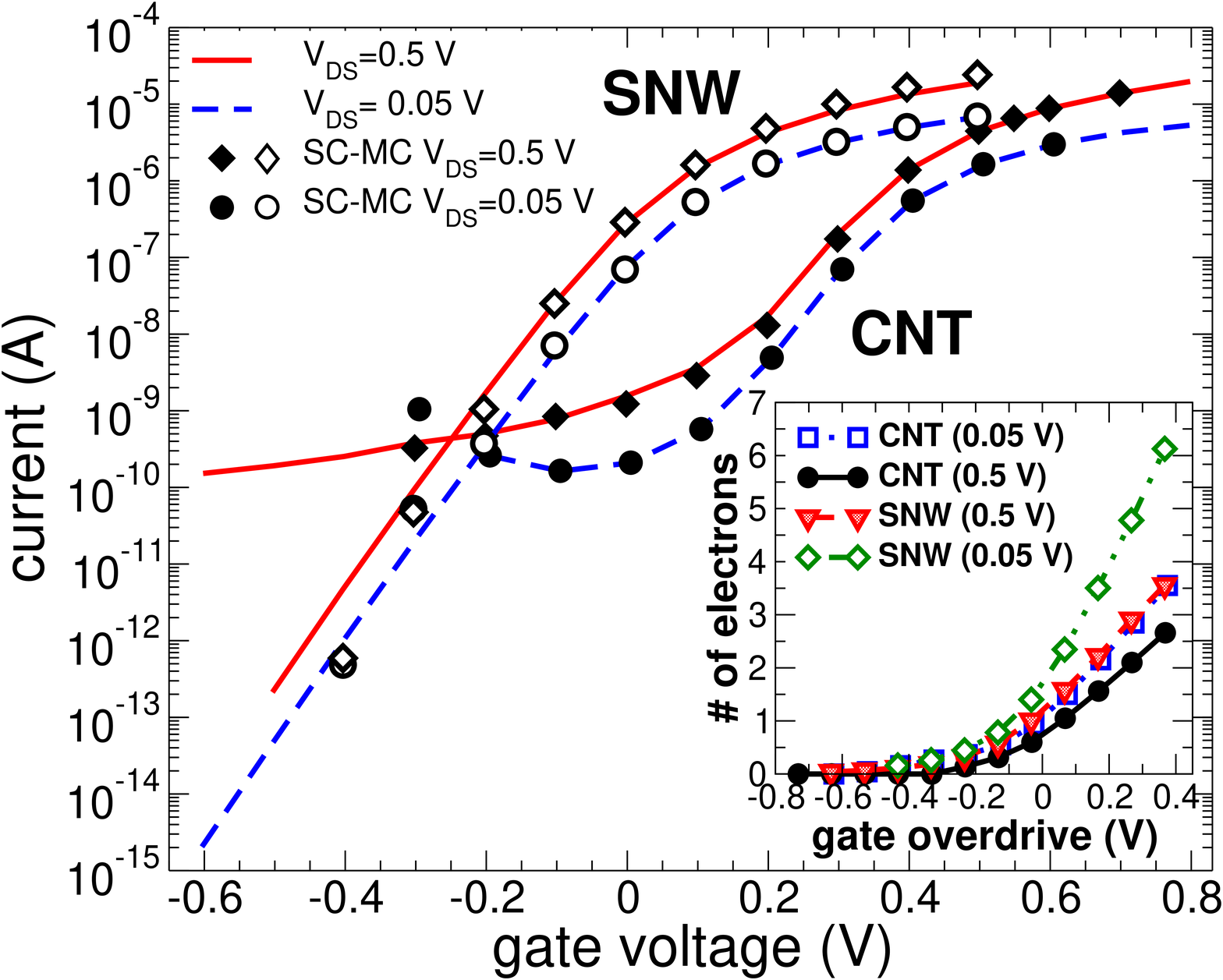}
\end{center}
\caption{Transfer characteristics computed for $ V_{DS}$= 0.5 V and 0.05 V, 
obtained by SC-MC and SC 
simulations, for CNT and SNW-FET. Full dots refer to CNT, empty dots to SNW. 
Inset: average number of electrons in CNT-FETs and SNW-FETs channel, evaluated for $ V_{DS}$= 0.5 V and 0.05 V.}
\label{fig:transferrandommedio}
\end{figure}

In Fig.~\ref{fig:transferrandommedio}, the transfer characteristics 
for different drain-to-source biases $V_{DS}$ computed performing SC and 
SC-MC simulations are plotted as a function of the gate 
overdrive $ V_{GS} - V_{th}$ 
in the logarithmic scale,  both for CNT and SNW devices. 
In particular the threshold voltage $V_{th}$ for the CNT-FET at $V_{DS}=$ 0.05 V and 0.5 V is 0.43 V, whereas we obtain $V_{th}=$ 0.13 V for $V_{DS}=$ 0.05 V and 0.5 V for the SNW-FET. 
As can be noted, SC and SC-MC simulations give practically the same results 
for CNT-FET, except in the subthreshold region where an interesting 
rectifying effect of the statistics emerges in the Monte Carlo simulations 
for a drain-to-source bias $V_{DS}=$~0.5~V. 

Instead, the rectifying effect is larger for SNW-FET, differences up to 
30 \% between the drain current 
$\langle I \rangle $ computed by means of SC-MC and SC simulations can be 
also observed in the above threshold regime. In particular, for a gate 
voltage $V_{GS}=$~0.5~V and a drain-to-source voltage $V_{DS}=$~0.5~V, 
the drain current $\langle I \rangle $ holds 
2.42 $\times$ 10$^{-5}$~A applying eq.~(\ref{eqn:meancurrent}), 
and 1.89 $\times$ 10$^{-5}$~A applying Landauer's 
formula~(\ref{eqn:Landauer}). 
Current in the CNT-FET transfer characteristics increases 
for negative gate voltages due to the interband tunneling. Indeed, the 
larger the negative gate voltage, the 
higher the number of electrons that tunnel from bound states in the 
valence band to the drain, leaving positive charge in the channel, which 
eventually lowers the barrier and increases the off current~\cite{GFiori}. 

In the inset of Fig.~\ref{fig:transferrandommedio} the average number 
of electrons inside the 
channel of a CNT and SNW-FET for two different biases $V_{DS}=$ 0.5 V and 
0.05 V is shown.
As can be seen, only very few electrons contribute to transport at any 
give instant, 
which requires us to attently evaluate the sensitivity of such devices to 
charge fluctuations: 
the smaller the drain-to-source voltage, the larger the average 
number of electrons in the channel, since, for low $V_{DS}$, carriers are 
injected from both contacts.  

\subsection{Noise}
\begin{figure} [tbp]
\begin{center}
\includegraphics[width=8.6cm]{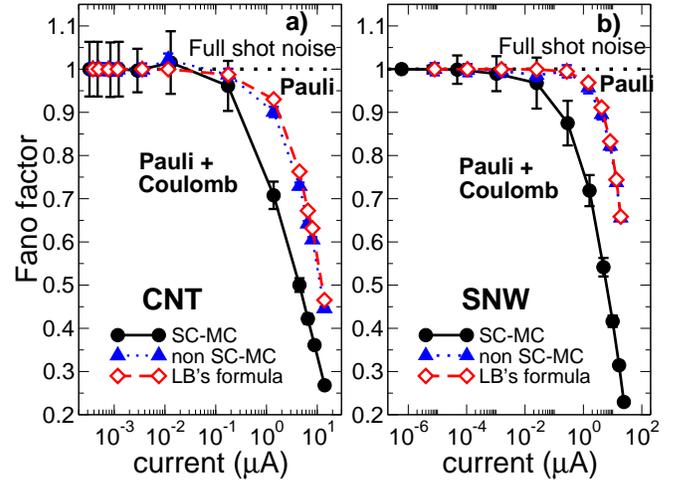}
\end{center}
\caption{Fano factor as a function of the drain current 
$\langle I \rangle$ for a) CNT- and b) SNW-FETs for $ V_{DS} = $ 0.5 V. 
Solid line refers to the Fano factor $F$ obtained by means of SC-MC 
simulations, dashed line (diamonds) applying eq. (\ref{eqn:noise}) and dotted 
line (triangles) by means of non SC-MC simulations.}
\label{fig:FanoCNTSNWvsI}
\end{figure}

Let us now focus our attention on the Fano factor $F$, defined as the ratio 
of the computed noise power 
spectral density $S(0)$ and the full shot noise $2 e \langle I \rangle $, 
$F=S(0)/(2 e \langle I \rangle)$. 
In Fig.~\ref{fig:FanoCNTSNWvsI}, the Fano factor 
for both CNT-FETs and SNW-FETs is shown for $ V_{DS}$= 0.5 V as a function 
of drain-to-source current $\langle I \rangle$. 

Let us discuss separately the effects of Pauli exclusion alone 
and concurrent Pauli and Coulomb interactions.
Triangles in Fig.~\ref{fig:FanoCNTSNWvsI} refer to $F$ computed 
by means of non SC-MC simulations on 
10$^4$ samples, while diamonds to results obtained by means of 
Landauer-B\"uttiker's formula, applying 
eq. (\ref{eqn:noise}). As expected the two approaches give the same 
results for both structures.
Solid lines refer to $S(0)$ computed by means of 
eqs.~(\ref{eqn:variance}) and~(\ref{eqn:noisepower}) and SC-MC simulations, 
i.e. Pauli and Coulomb interactions simultaneously taken into account.

In the sub-threshold regime 
($\langle I \rangle<$ 10$^{-9}$ A), 
drain current noise is very close to the full shot noise, since 
electron-electron
correlations are negligible due to the very small amount of mobile charge 
in the channel.

From the point of view of eq.~(\ref{eqn:noise}), 
for energies larger than the top of the barrier, we have 
$f_D(E)\ll f_S(E)\ll \,1$ and the 
integrand in~(\ref{eqn:noise}) reduces to $\mathrm{Tr}\left[\mathbf{t^\dagger t}(E)\right]f_S(E)$. Instead, for energies smaller than the high potential 
profile 
along the channel, $\left[\mathbf{t^{\dagger} t}(E)\right]_{n m} \ll 1$ $\forall n,\, m \in S$, so that we can neglect $\mathrm{Tr}\left[\mathbf{t^\dagger t t^\dagger t} \right]$ in~(\ref{eqn:noise}), with respect to $\mathrm{Tr}\left[\mathbf{t^\dagger t} \right]$. Since $f_D(E) \ll f_S(E)$, the integrand in (\ref{eqn:noise}) still reduces to $\mathrm{Tr}\left[\mathbf{t^\dagger t}(E)\right]f_S(E)$. 
The Fano factor then becomes
\begin{eqnarray} \label{eqn:fullshot}
F = \frac{S(0)}{2 e \langle I \rangle} \approx \frac{ \frac{2\,e^2}{\pi\hbar} \int d\!E \,  \mathrm{Tr}\left[\mathbf{t^\dagger t}(E) \right] f_S(E) }
{ 2 e \frac{e}{\pi\hbar} \int d\!E \, \mathrm{Tr}\left[\mathbf{t^\dagger t}(E)\right] f_S(E) }=\, 1
\end{eqnarray} 

In the strong inversion regime instead 
($\langle I \rangle>$ 10$^{-6}$ A), the noise is 
strongly suppressed with respect to the full shot value.
In particular for a SNW-FET, at $\langle I \rangle\approx$ 2.4 $\times$ 10$^{-5}$~A ($ V_{GS} - V_{th}\approx$ 0.4 V), combined Pauli and Coulomb interactions suppress shot 
noise down to 23 \% of the full shot noise value, with a significant 
reduction with respect to the value predicted without 
including space charge effects as in Ref.~\cite{Park1}, while for CNT-FET the 
Fano factor is equal to 0.27 at 
$\langle I \rangle\approx$ 1.4 $\times$ 10$^{-5}$~A 
($ V_{GS} - V_{th}\approx$ 0.3~V). 
Indeed, an injected electron tends to increase 
the potential barrier along 
the channel, leading to a reduction of 
the space charge and to a suppression of charge fluctuation.
Note that, by only considering Pauli exclusion principle, we would 
overestimate 
shot noise by 180 \% for SNWT 
($\langle I \rangle\approx$ 2.4 $\times$ 10$^{-5}$ A) and 
by 70 \% for CNT-FET ($\langle I \rangle\approx$ 1.4 $\times$ 10$^{-5}$ A). 

\subsection{Shot noise versus thermal channel noise}
According to the classical approach for the formulation of drain current 
noise, channel noise is tipically 
described in terms of a ``modified'' thermal noise, as
$S(0) = \gamma S_T$, where $S_T=4K_BT g_{d0}$ is the thermal noise power 
spectrum at zero drain-to-source bias $V_{DS}$, $k_B$ is the Boltzmann 
constant, $\gamma$ is a correction parameter and 
$g_{d0}= \, \left(\partial \langle I \rangle/\partial V_{DS}\right)_{V_{DS}= \, 0} $ is the 
source-to-drain conductance at zero $V_{DS}$. 

Although the classical formulation accurately predicts drain current noise 
in long channel MOSFETs, where $\gamma$ is equal to 1 in the ohmic 
region and 2/3 in 
saturation~\cite{Ziel}, it underestimates noise in short channel devices. 
In particular, experimental evidences~\cite{Abidi} of an excess noise in 
short channel 
MOSFET have been explained in terms of the limited number of scattering events 
inside the channel which is uneffective in suppressing the non-equilibrium 
noise component~\cite{RNavid}, or in terms of a revised classical 
formulation by considering 
short channel effects, such as the carrier heating effect above the lattice 
temperature~\cite{KHan}. 

\begin{figure} [tbp]
\begin{center}
\includegraphics[width=8.6cm]{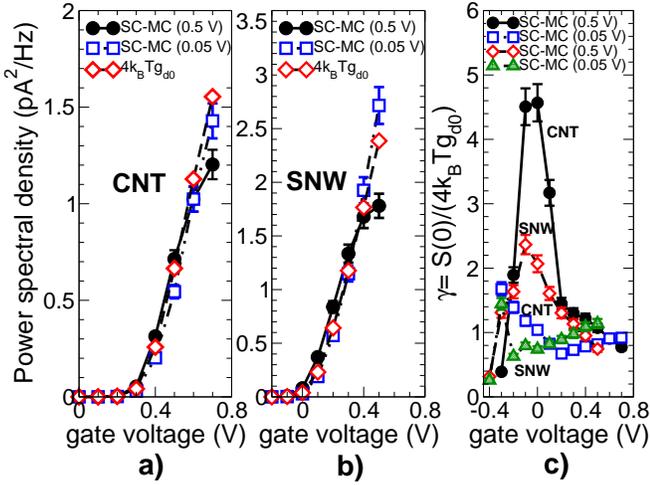}
\end{center}
\caption{(a) noise power spectral density obtained by SC-MC simulations and thermal noise spectral density as functions of the gate voltage for a) CNT-FETs and b) SNW-FETs: 
the considered drain-to-source biases ($ V_{DS}=$ 0.5 V, 0.05 V) are shown in brackets; c) ratio between the noise power obtained by SC-MC simulations and the thermal noise density as a function of the gate voltage. $ g_{d0} $ is the conductance evaluated for $V_{DS}=$ 0 V: $g_{d0}=\, \left(\frac{\partial \langle I \rangle}{\partial V_{DS}}\right)_{V_{DS}=0} $.}
\label{fig:thermnoise1}
\end{figure}

Actually, it can clearly be seen that non equilibrium transport easily 
provides $\gamma > 1$ and that the cause of $\gamma > 1$ is simply due to 
the fact that channel noise can be more properly interpreted as shot noise. 
For example, in the particular case of ballistic transport considered here, 
we can plot $\gamma$ as $S(0)/ S_T$ as a function of the gate voltage in 
Fig.~\ref{fig:thermnoise1}c. As can be seen, values of $\gamma$ larger 
than 1 can be easily observed in weak and strong inversion. 
The strange behavior of $\gamma$ as a function of the gate voltage is 
simply due to the fact that one uses an inadequate model (thermal noise) 
corrected with the $\gamma$
parameter to describe a qualitatively different type of noise, 
i.e. shot noise. 
\begin{figure} [tbp]
\begin{center}
\includegraphics[width=8.6cm]{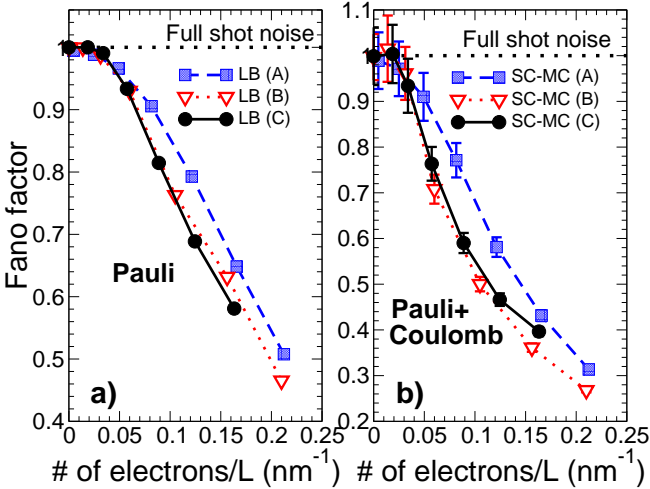}
\end{center}
\caption{Fano factor as a function of the average number of electrons 
inside the channel per unit length 
for three different (13,0) CNT-FETs: (A) $t_{ox}$= 1 nm, $L$= 6 nm, 
(B) $t_{ox}$= 1 nm, $L$= 10 nm and (C) $t_{ox}$= 2 nm, $L$= 10 nm. 
In a) only the effect of the Pauli principle is shown [eq.~(\ref{eqn:noise})]; 
in b) the effect of both Pauli and Coulomb interactions 
is considered. The drain-to-source bias $V_{DS}$ is 0.5 V.}
\label{fig:Fscaling}
\end{figure}

\subsection{Effect of scaling on noise}
Let us now discuss the effect of scaling on noise, focusing our 
attention on a (13,0) CNT-FET. 
One would expect that an increase of the oxide thickness would reduce the 
screening induced by the metallic gate, so that the 
Coulomb interaction would be expected to produce a larger noise suppression. 
For example, in the limit of a multimode ballistic conductor without a 
gate contact, significantly suppression of about two 
order of magnitude with respect to the full shot value has been shown 
by Bulashenko et al~\cite{Bul1}. 

However, Ref.~\cite{Bul1} exploits a semiclassical approach assuming a large 
number of modes and the conservation of transversal momentum, i.e. the role of 
the transversal electric field induced by the gate voltage is completely 
neglected. 
In our case only four modes contribute to transport, while 
the top and bottom gates of 
the simulated devices partially screen the electrostatic repulsion induced 
by the space charge in the channel on each injected electron, so that a 
smaller noise suppression than the one achieved in Ref.~\cite{Bul1} can be 
expected. 

The Fano factor as a function of the average number of electrons inside the 
channel for unit length, computed by means of SC simulation and applying 
eq.~(\ref{eqn:noise}),
for three CNTs with different oxide thickness $t_{ox}$ and 
channel length $L$ is shown in Fig.~\ref{fig:Fscaling}a: it shows results 
for CNT with $t_{ox}$= 1 nm, $L$= 6 nm (A), CNT with $t_{ox}$= 1 nm, 
$L$= 10 nm (B), and CNT with $t_{ox}$= 2 nm, $L$= 10 nm (C).
Fig.~\ref{fig:Fscaling}b shows the Fano factor computed by performing 
SC-MC simulations and applying eqs.~(\ref{eqn:variance}) and 
(\ref{eqn:noisepower}). 
As can be seen, if the Fano factor is plotted as a function of the number 
of electrons per unit length, as in Fig.~\ref{fig:Fscaling}, curves are 
very close to one another, and effects of scaling are predictable. 

\section{Conclusion}
We have developed a novel and general approach to study 
shot noise in nanoscale quasi one-dimensional FETs, such as CNT-FETs and 
SNW-FETs. Our first important result is the derivation of an analytical 
formula for the noise power spectral density which exploits a statistical 
approach 
and the second quantization formalism. Our formula extends the validity of the 
Landauer-Buttiker noise formula [eq.~(\ref{eqn:noise})], to 
include also Coulomb repulsion among electrons. From a quantitative point of 
view, this is very important,
since we show that by only using Landauer-Buttiker noise formula, one can 
overestimate shot noise by as much as 180\%. 
The second important result is the implementation of the method in a 
computational code,
based on the 3D self-consistent solution of Poisson and Schr\"odinger 
equation with the 
NEGF formalism, and on Monte Carlo simulations over a large ensemble 
of occurrencies, with random
occupation of electronic states incoming from the reservoirs. 
As a final note, we show that scaling of ballistic onedimensional FETs 
is expected to weakly affect drain current fluctuations, even in 
the degenerate injection limit.

\section{Acknowledgment}
The work was supported in part by the EC Seventh 
Framework Program under the Network of Excellence NANOSIL (Contract 216171), 
and by the European Science Foundation EUROCORES Program Fundamentals of 
Nanoelectronics, through funds from CNR and the EC Sixth Framework Program, 
under project DEWINT (ContractERAS-CT-2003-980409).



\end{document}